\documentclass{amia}
\usepackage{graphicx}
\usepackage[labelfont=bf]{caption}
\usepackage[superscript,nomove]{cite}
\usepackage{color}
\usepackage[linesnumbered,ruled,vlined]{algorithm2e}
\usepackage{amsmath}
\usepackage{subcaption}
\usepackage{url}
\usepackage{multirow}
\usepackage{makecell}
\usepackage{rotating}
\begin{document}

\title{A Systematic Approach to Surveillance and Detection of Hierarchical Healthcare Cost Drivers and Utilization Offsets}

\author{Ta-Hsin Li, PhD$^{1}$, Huijing Jiang, PhD$^{1}$, Kevin Tran, PhD$^{2}$, Gigi Yuen-Reed, PhD$^{2}$, Bob Kelley, PhD$^{2}$, Thomas Halvorson, BS$^{2}$}

\institutes{
    $^1$IBM T.J. Watson Research Center, Yorktown Heights, NY, USA; $^2$IBM Watson Health, Cambridge, MA, USA\\
}

\maketitle

\noindent{\bf Abstract}

\textit{There is strong interest among healthcare payers to identify emerging healthcare cost drivers to support early intervention. However, many challenges arise in analyzing large, high dimensional, and noisy healthcare data. In this paper, we propose a systematic approach that utilizes hierarchical search strategies and enhanced statistical process control (SPC) algorithms to surface high impact cost drivers. Our approach aims to provide interpretable, detailed, and actionable insights of detected change patterns attributing to multiple clinical factors. We also proposed an algorithm to identify comparable treatment offsets at the population level and quantify the cost impact on their utilization changes. To illustrate our approach, we apply it to the IBM Watson Health MarketScan Commercial Database and organized the detected emerging drivers into 5 categories for reporting. We also discuss some findings in this analysis and potential actions in mitigating the impact of the drivers.}

\section*{Introduction}
There is strong interest to better understand and manage drivers of healthcare cost. Payers, including public agencies, private health plans, and self-insured employers, are particularly interested in identifying emerging cost drivers to support early intervention. Traditional approaches to identifying cost drivers within large payer claims databases can be labor-intensive. It may require manually drilling into the data, and often times, drivers exhibit insidious trends, are masked within summary reports, or are too general to deem actionable.

Data scientists tasked with cost driver detection can be overwhelmed by the sheer volume of data. Their analysis may be driven by personal experience which could bias their approach to perform a systematic and comprehensive detection. Furthermore, several factors make analyzing healthcare claims data a challenge. First, claims data is high dimensional and sparse with noisy signal strength. Second, data collection and standardization can be inconsistent across data sources, due to variation in clinical coding practices, healthcare delivery and payment models, and speed of claims processing. Third, healthcare cost drivers tend to be inter-related, making it difficult to isolate the underlying root cause. Finally, healthcare cost changes could be influenced by seasonal factors, clinical guidelines, or new technologies introduced to the market. These considerations make it challenging for data scientists and payers to determine where to efficiently focus their efforts on.

In this paper, we introduce a systematic approach to surveillance and detection of emerging healthcare cost drivers. It combines the statistical process control (SPC) method with the hierarchy of healthcare services and other contributing factors to cost change. 

Applications to SPC have been used in many healthcare domains\cite{keller2015, ray2017, thor2007}. However, existing studies have been limited to a specific outcome within a target population (e.g., inpatient readmission rates for an individual hospital). Instead of looking at a particular condition or treatment, our approach performs a large scale search across hierarchical paths. We utilize hierarchical schemes to surface the emerging cost drivers and attribute the cost changes to multiple explanatory factors. Compared to other cost attribution studies\cite{katz2015}, we leverage clinical episode-based groupings and multi-factor drill downs for better interpretability. We further evaluate the impact of cost change based on a hierarchy of contributing factors including price and use of treatments and prevalence of medical conditions. We also propose an algorithm to monitor ``utilization offsets'' of comparable treatments and estimate the cost impact attributed to the offset effect. Utilization offsets or treatment switch analyses have been studied extensively within real-world evidence applications\cite{feldman2019}. Our approach extends this concept to evaluate the cost impact of offsets for drug treatments, disease severity, as well as care setting. The proposed surveillance and detection system should be executed regularly and frequently over time based on the business requirements (e.g., once every month, quarter, or year) in order to identify emerging cost drives in a timely manner. 

We illustrate our approach through a use case study on IBM Watson Health MarketScan Commercial Database. The emerged cost drivers are organized into 5 catogories for better reporting. 

\section*{Methods}
\subsection*{Search Strategy}
We apply a hierarchical search strategy leveraging domain knowledge-based factors to search for impactful cost drivers. The claims data is aggregated into multiple key performance indicator (KPI) time series based on different hierarchical drill paths or ``viewpoints''.

We use a 1 million random sample of enrollees from the IBM Watson Health MarketScan Commercial Database \cite{marketscan} from 2012-2016. The database includes a longitudinal perspective of enrollment, demographic, medical and pharmacy claims data from large employers and health plans. The claims data provides demographic and clinical information (e.g., diagnoses, procedures, drug codes, care setting, etc.) and cost information associated with the healthcare services. We also utilize medical groupers that combine medical and pharmacy claims into unique episodes of care which allows us to tie drug information to specific conditions \cite{MEG}. Finally, we use the Micromedex RED BOOK database to establish a knowledge base of comparable drug treatments for the offsets identification algorithm\cite{redbook}.

\begin{figure}[h!]
	\centering
	\begin{subfigure}{0.52\textwidth}
		\centering
		\includegraphics[width=\linewidth]{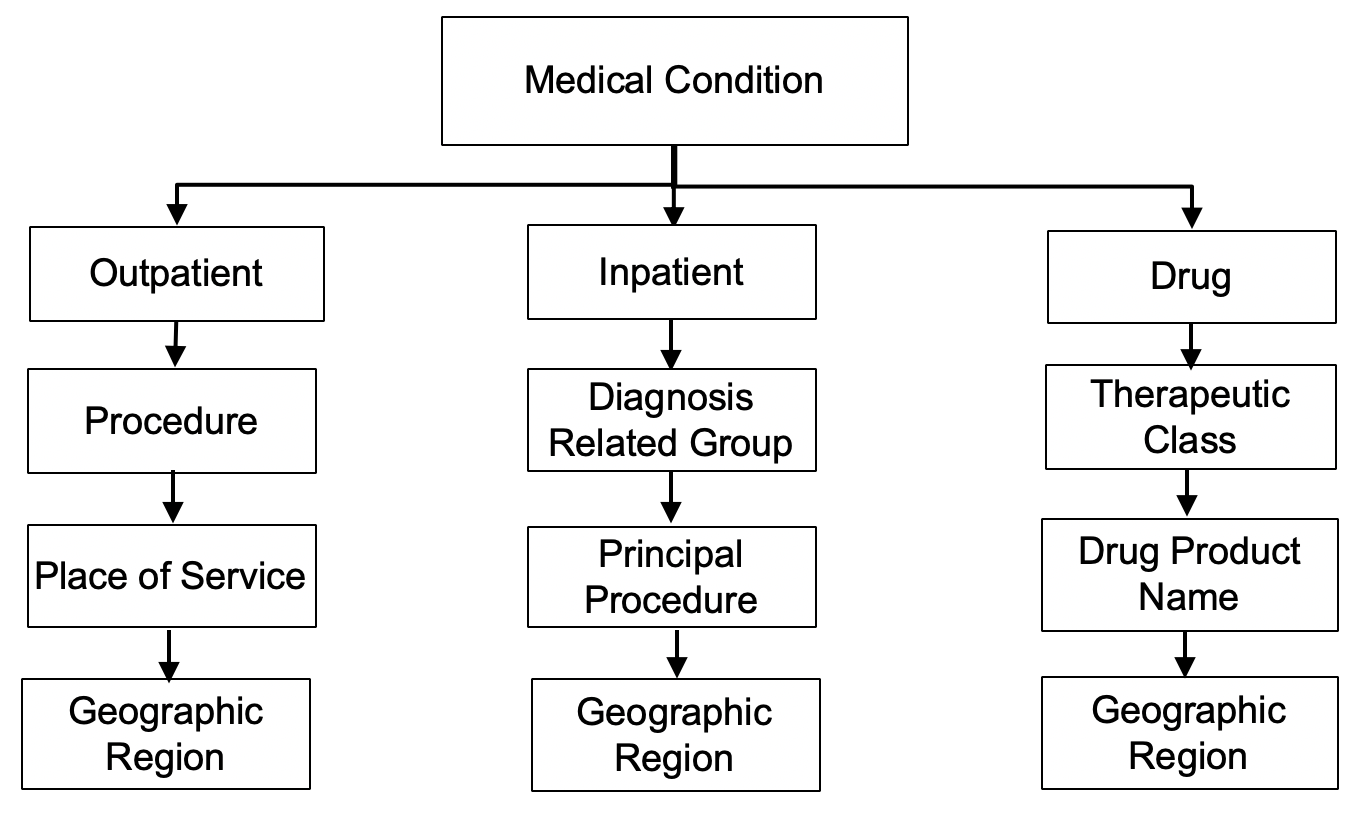}
		\caption{}
		\label{fig:hierarchy:condition}			
	\end{subfigure}
	\begin{subfigure}{0.2\textwidth}
		\vspace{2\baselineskip}
		\centering
		\includegraphics[width=\linewidth]{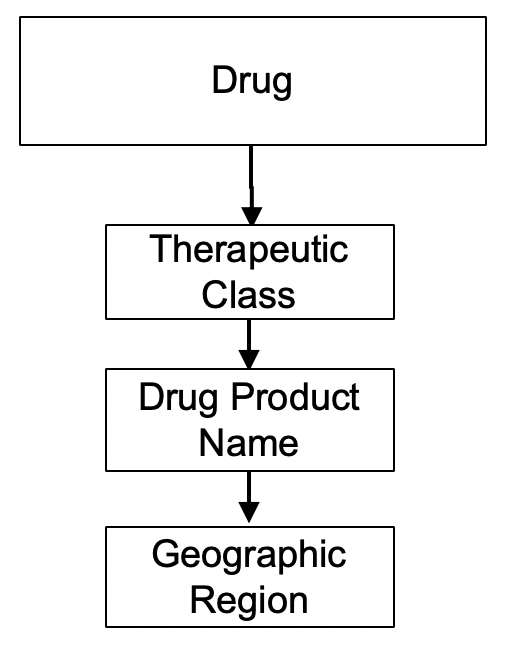}	
		\caption{}
		\label{fig:hierarchy:drug}	
	\end{subfigure}
	\caption{Example of Viewpoint Hierarchy.}	
	\label{fig:hierarchy}	
\end{figure}

The claims are  aggregated according to viewpoint definitions. Figure \ref{fig:hierarchy:condition} shows an example of different viewpoints based on clinical attributes grouped into clinical episodes of care by assigning an event label (e.g., treatment episodes, admissions) to each claim record using IBM Watson Health Analytics Engine (AE). For example, a viewpoint pertaining to a specific condition, pharmacy claim type, therapeutic class, and drug product name hierarchy will provide insights on a specific drug for treating a specific condition. This can also be extended to inpatient and outpatient claims and their related attributes (e.g., procedure, place of service, geographical region, etc.). It is also possible to start the hierarchy without medical conditions to focus on treatment costs in general. For example, drug costs in general can be monitored using the hierarchy shown in Figure \ref{fig:hierarchy:drug}.


Various KPIs are computed for the time series. For each viewpoint, we calculate the total cost, number of episodes, number of enrollees, number of patients with a specific condition, number of claimants on a specific treatment, and quantity of services. This provides information to calculate the ratios for each KPI and standard errors.

\subsection*{Change Patterns Detection}\label{sec:detect}
To detect changes early, reliably, and in a clinically meaningful manner, we developed an enhanced SPC-based analytics algorithm incorporating domain hierarchical knowledge. It contains three primary functionalities: a threshold learning module to establish a detection threshold through historical data modeling, an online detection mechanism (e.g., auto-reset or non-restarting CUSUM \cite{gandy2013, lau2013}), and a change pattern reporting rules engine.

The enhanced SPC-based algorithms use one or more detection thresholds to control the false detection rate. Detection thresholds are learned via simulation to account for sampling errors and possible serial dependence of change rates. For example, statistical time series models (e.g., ARMA, Gaussian white noise, etc.) for normalized rates of change could be used to simulate time series data for different KPIs under the hypothesis of no change. We run the SPC-based change detection algorithms over each simulated time series for each value in the set of trail thresholds. The fraction of detected cases (false alarm rate) are then computed for each trial threshold value. The threshold whose false alarm rate is closest to the target false alarm rate is identified.

\begin{figure}[h!]	
	\centering	
	\includegraphics[width=.8\linewidth]{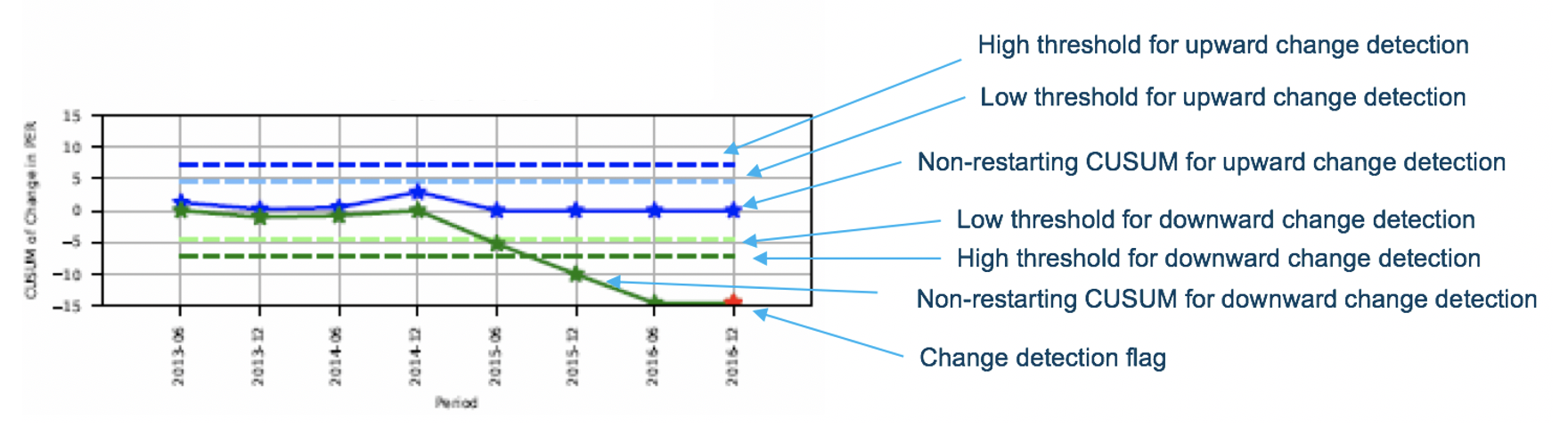}
	\caption{Example of Enhanced SPC Algorithms: Non-restarting CUSUM}
	\label{fig:cusum}
\end{figure}

The online detection algorithms automatically account for multiple change points. By specifying reporting rules, detected changes can be reported at any point within the analysis window, or reported at the end of the analysis to focus on the latest cumulative effect of change. Figure \ref{fig:cusum} shows an example of enhanced SPC algorithms with learned detection thresholds. High and low thresholds for upward and downward change detection are learned via simulation. In this example, the focus is on the latest cumulative effect of change. The non-restarting CUSUMs is applied here and the change detection flag is only triggered at the end of the analysis time period.

The impact of change on the average expenditure per enrollee is determined by
$$c(t)=s(t)-s(t-T), t=T, \ldots, P$$
$$I(c)=EWA(c(T+1), \ldots, c(P))$$
where $s(t)$ is the cost per enrollee in period $t$, and EWA is the exponential weighted average $\frac{1-w}{1-w^{P-T}}\sum^{P}_{t=T+1}w^{P-t}c(t)$, where the weight $w$ is a number between 0 and 1 which enables to downplay the changes in older periods and emphasize the changes in more recent ones. One could also replace the weighted average by a simple average where changes in all periods play an equal role.

\begin{figure}[h!]	
	\centering	
	\includegraphics[width=.8\linewidth]{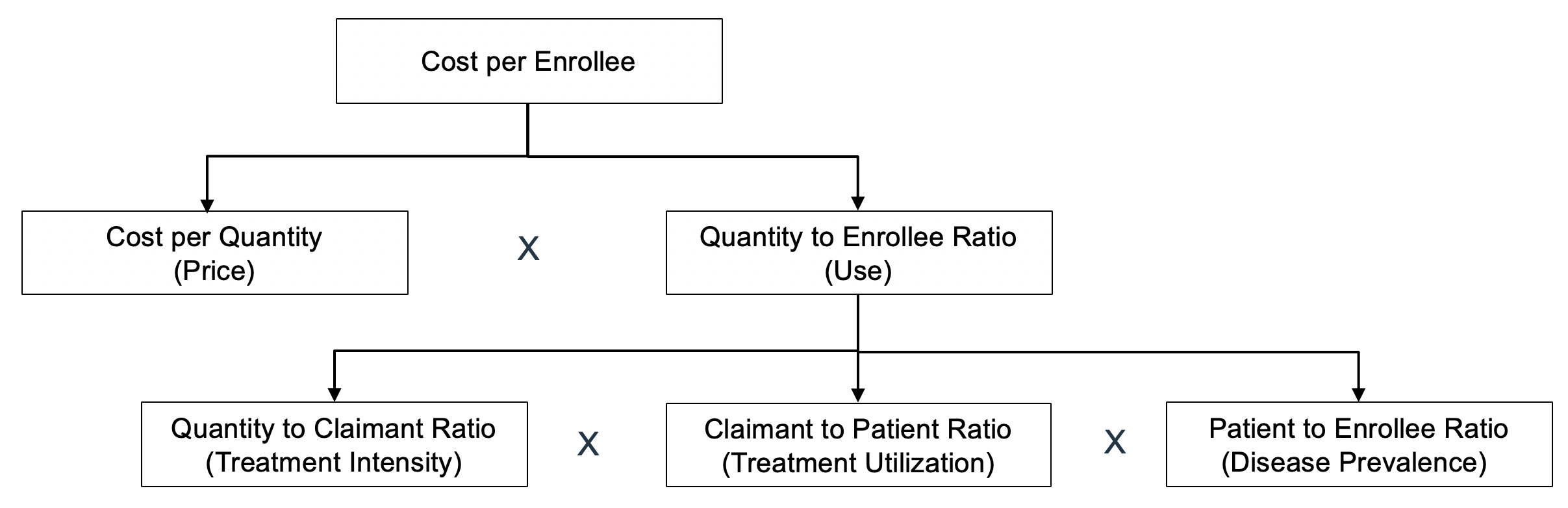}	
	\caption{Example of Multi-KPI Change Attribution}
	\label{fig:decomposition}
\end{figure}

The total impact of change $I(c)$ can be further decomposed into impact from contributing factors according the hierarchy shown in Figure 3. For example, $I(c)$ can be decomposed into contributions from price change, $I(c_1)$, and from use change, $I(c_2)$, such that $I(c)=I(c_1)+I(c_2)$. This is accomplished by allowing the factor of interest to change into new values while holding the other factors at past values. For the price and use decomposition, it can done as follows:
\begin{align*}
c_1(t)&=e(t-T)[a(t)-a(t-T)], t=T+1, \ldots, P\\
J(c_1)&=EWA(c_1(T+1), \ldots, c_1(P))\\
c_2(t)&=[e(t)-e(t-T)]a(t-T), t=T+1, \ldots, P\\
J(c_2)&=EWA(c_2(T+1), \ldots, c_2(P))
\end{align*}
where $e(t)$ is the use in period $t$, $a(t)$ is the price in period $t$. Let $\delta_1=[J(c_1)+J(c_2)]-I(c)$. Then, the impact due to price is defined as
$$I(c_1)=J(c_1)-\delta_1\times|J(c_1)|/[|J(c_1)|+|J(c_2)|],$$
and the impact of change due to use is defined as
$$I(c_2)=J(c_2)-\delta_1\times|J(c_2)|/[|J(c_1)|+|J(c_2)|].$$

In the last two equations, the preliminary contributions $J(c_1)$ and $J(c_2)$ are adjusted proportionally by the remainder of preliminary decomposition $\delta_1$ to make the sum of the adjusted contributions equal to the total impact $I(c)$. Similarly, as illustrated by Figure \ref{fig:decomposition}, the impact of change due to use $I(c_2)$ could be further decomposed by treatment intensity, treatment utilization and disease prevalence.

\subsection*{Utilization Offsets Identification}

The offsets are identified when utilization patterns of the comparable treatments under the same condition move in opposite directions within the same time window. Offsets may include drug treatments, place of services etc. Potential substitutions of drugs within the same therapeutic class and indication per published evidence may be analyzed using Micromedex RED BOOK database. Shifts in care settings may also be identified (e.g., ambulatory surgical center vs hostipal for screening colonoscopy). The method can also be extended to identify the progression of disease severity stage based on the changes of disease prevalence patterns.

Offset tracking in general can be accomplished at different granularities, from population-level to individual-level. Our approach focuses on estimating the impact using population-level data which reduces the analytic complexity. We define an offsetting network including originators (treatments with decreased utilization) and receivers (treatments with increased utilization). 
The utilization migration within the offsetting network and external factors are difficult to track at the population level; we cannot distinguish individuals who are existing versus new or dropout for a treatment. We only have information on the total usage of the treatment at sub-population level, which poses challenges to accurately estimate the utilization migration. To determine the volume of utilization migration excluding the effects of new/dropout patients, we postulate the following assumptions.

{\itshape Proportional Allocation Assumptions.} (a) The volume of outflow offset from a treatment option that experiences utilization decrease (originator) to a comparable treatment option that experiences utilization increase (receiver) is assumed to be proportional to the amount of observed utilization increase of the receiver. (b) The total volume of outflow offset from an originator is assumed to be proportional to the amount of observed utilization decrease of the originator.

{\itshape Migration Equilibrium Equations.} The offset inflow of a receiver is equal to the total offset outflow received from all of its originators.

{\itshape Maximum Migration Principle.} The outflow offset volumes from originators to receivers are determined by maximizing the total amount of outflow offset from all originators under the constraints of proportional allocation. The inflow offset volume of each receiver is determined by summing the outflow volumes from all originators.

Utilization migration for each treatment options within the network can be calculated based on Algorithm \ref{algo:MODimpact}. The cost impact due to offset can be computed by using offset-adjusted utilization under the assumption of no change in average cost. 

\begin{algorithm}
	
	\DontPrintSemicolon 
	
	\KwIn{An offseting network with outflow $o_1,\ldots, o_I$ from the originators $O_1, \ldots, O_I$ and inflow $r_1, \ldots, r_J$ to receivers $R_1, \ldots, R_J$ .}
	
	\KwOut{Offset outflow $o_{m,1},\ldots, o_{m,I}$ and offset inflow $r_{m,1},\ldots, r_{m,I}$.}
	
	Estimate the population of migrations that satisfies Maximum Migration Principle and the migration constrains	
	$P_m = \min\left[\sum_{i=1}^{I}o_i, 1/\left(\sum_{i=1}^{I}\frac{o_i}{\sum R(o_i)\sum_{i=1}^{I}o_i}\right) \right] $\\ 
	
	Given migration population $P_m$, calculate within-network migration from each originators $o_{m, 1}, \ldots, o_{m, I}$ based on Proportional Allocation assumptions (b).\\
	
	Calculate migration transitions $(O_i \rightarrow R(O_i))$ based on Proportional Allocation assumptions (a).\\
	
	Calculate migration to each receivers $r_{m,1},\ldots, r_{m,I}$ based on Migration Equilibrium Equations.
	
	\caption{Algorithm of Calculating MOD Impact}
	
	\label{algo:MODimpact}
	
\end{algorithm}

\section*{Results}
In this study, we run our analysis yearly on MarketScan data from 2013 to 2016. In practice, we can get early-warning alerts if we run the analysis monthly instead of yearly. We divide and report all detected emerging cost drivers into 5 categories: (1) Pharmacy Drug, (2) Condition Management, (3) System Improvement, (4) Rare Diseases and (5) Declining Cost.  In this session, we deep dive into some examples in the first three categories. 

{\bf  Pharmacy Drug}

Drug category analyzes cost drivers related to prescription drugs. In this category, Glumetza shows up as one of the top drivers. It is a diabetes pill produced by Salix Pharmaceuticals before Salix was acquired by Valeant. On June 18, 2015, Valeant raised the price of Glumetza from \$572 to \$3,432. About one month later, Valent hiked Glumetza's price again to \$5,148 \cite{nyt2015, nyt2016}. As shown in Table \ref{tab:glumetza_impact}, among about 850 pharmacy drugs in our study, the total cost of Glumetza skyrocketed from 0.02\% in 2014 to 0.06\% in 2015. The rate of cost change is as high as 262\% and the rank of cost change impact jumped from 179th in 2014 to 75th in 2015. 

\begin{figure}[h]
	\begin{subfigure}{.24\textwidth}
		\centering
		\includegraphics[width=\linewidth]{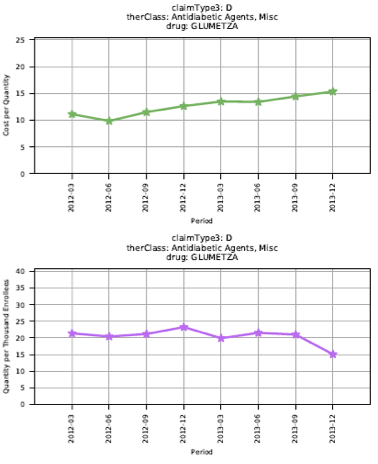}	
		\caption{12/2013 run}		
	\end{subfigure}
	\begin{subfigure}{.24\textwidth}
		\centering
		\includegraphics[width=\linewidth]{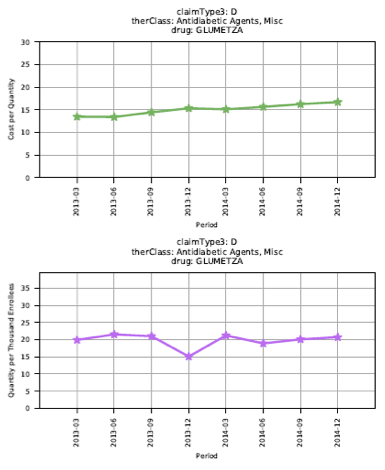}	
		\caption{12/2014 run}	
	\end{subfigure}
\begin{subfigure}{.24\textwidth}
	\centering
	\includegraphics[width=\linewidth]{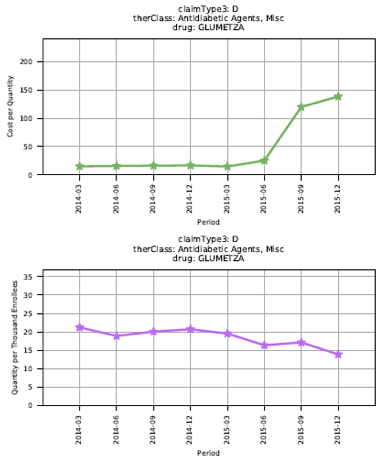}	
	\caption{12/2015 run}		
\end{subfigure}
\begin{subfigure}{.24\textwidth}
	\centering
	\includegraphics[width=\linewidth]{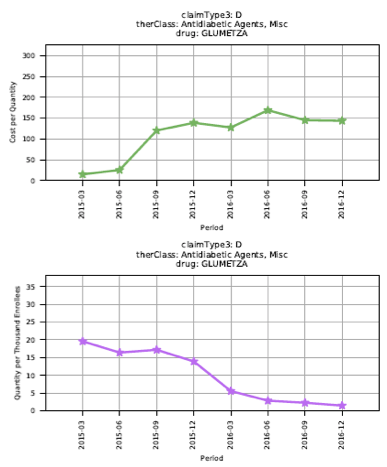}	
	\caption{12/2016 run}	
\end{subfigure}
	\caption{Price and use trends of Glumetza from 2013 to 2016.}	
	\label{fig:glumetza_trends}	
\end{figure}

A deep analysis using our proposed method shows that this dramatical price hike significantly impacted diabetes drug market.  As shown in Figure \ref{fig:glumetza_trends}, although there is already a strong evidence of price increase for Glumetza in 2014, there is no sufficient evidence of change in the use of Glumetza. However,  the use of Glumetza started to show a decreasing trend in 2015. Eventually, the use of Glumetza phased out in 2016. Actually, insurance companies started to ban Glumetza from their coverage list in order to tamp down drug prices. For example, in February 2016, Glumetza was excluded by Express Scripts, a pharmacy benefit management (PBM) giant\cite{fp2016}. 


Table \ref{tab:glumetza_impact} shows a decomposition of the total cost impact of this detected change into price (average cost per day of supply) and use (average number of days of supply per enrollee). The price has a very strong increasing trend in 2015. The rate of change is as high as 369\% and the rank of its impact jumped to the 3rd most positvely impactful drugs. On the other hand, the cost impact attributable to use has a moderate decreasing trend in 2015 and an even more stronger decreasing trend in 2016. As a result,  by the end of 2016, although the cost impact of price remains positive, the total cost impact became negative because of the dramatic decrease of its usage.



\begin{figure}[h]
	\begin{subfigure}{.5\textwidth}
		\centering
		\includegraphics[width=\linewidth]{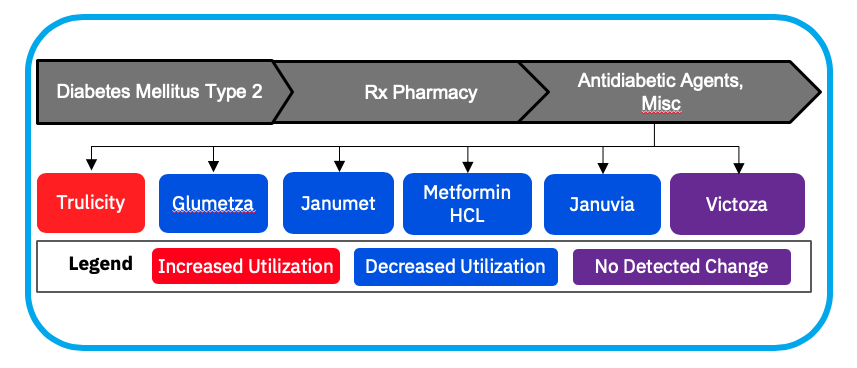}	
		\caption{Offsetting drugs of Glumetza and their trending directions.}	
		\label{fig:glumetza_mod}	
	\end{subfigure}
	\begin{subfigure}{.5\textwidth}
		\centering
		\includegraphics[width=\linewidth]{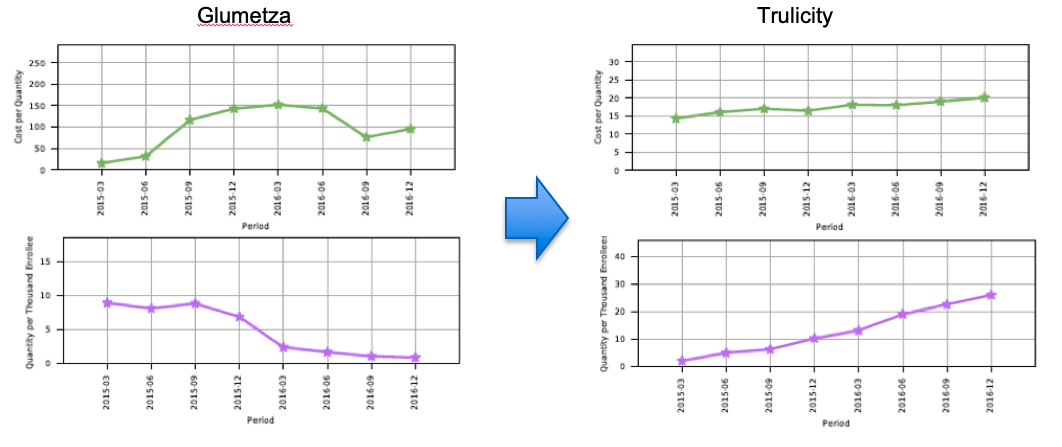}	
		\caption{Use flow from Glumetza to Trulicity.}	
		\label{fig:glumetza_trulicity}	
	\end{subfigure}
	\caption{Impact of Gluemtza's price change to the diabete drug market.}	
\end{figure}

Considering the high price of Glumetza, healthcare providers and patents started switching to alternative drugs. Figure \ref{fig:glumetza_mod} lists the offsetting drugs of Glumetza. All of them share the same generic formula, i.e., Metformin Hydrochloride as Glumetza. Among these offsetting drugs, Figure \ref{fig:glumetza_trulicity} shows the use of Trulicity has been persistently increasing ever since June 2016, the month when Glumetza suddenly hiked up its price. According to our MOD impact calculation, Trulicity substituting Metformin Hydrochlorid-based drugs such as Glumetza resulted in a cost decrease of \$0.38 PMPM for treating type 2 diabetes.

{\bf Condition Management}

Condition management category monitors the cost and prevalence of different condition drivers. Coxsackie and ECHO Infections turns out to be one of the conditions that experienced dramatic increase in prevalence.  

\begin{figure}[h!]
	\centering
	\includegraphics[width=0.6\linewidth]{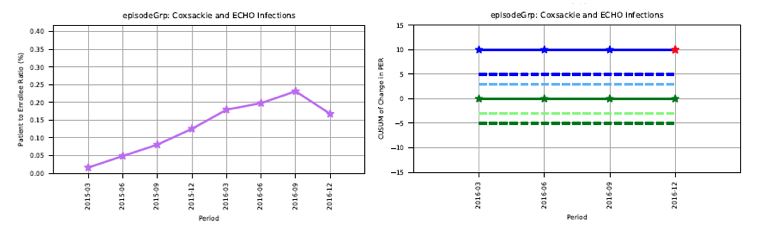}	
	\caption{Prevalance trends and CUMUM plots of Coxsackie and ECHO Infections in 2016}	
	\label{fig:coxs_trends}	
\end{figure}

\begin{figure}[h!]
	\centering
	\includegraphics[width=0.6\linewidth]{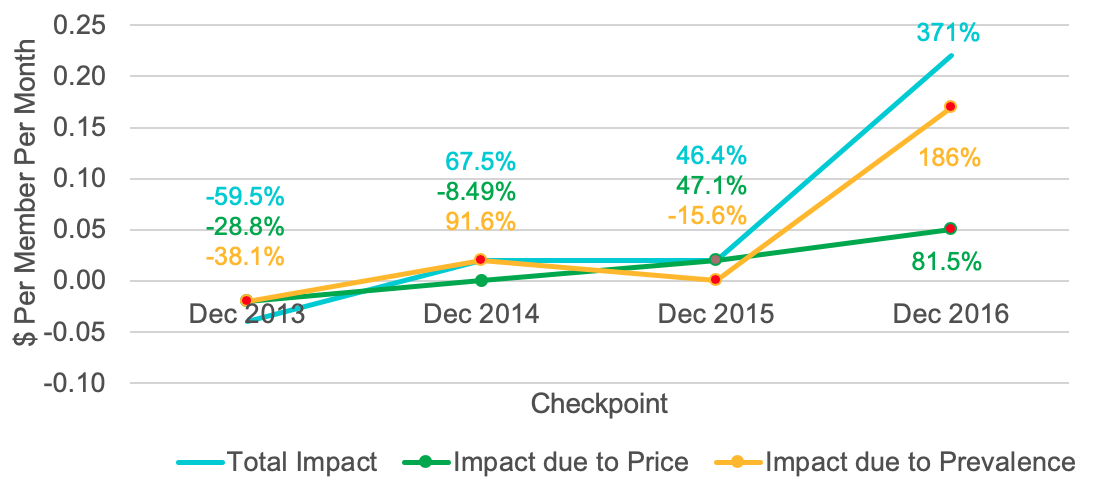}	
	\caption{Impact changes of Coxsackie and ECHO Infections from 2013 to 2016}	
	\label{fig:coxs_impact}	
\end{figure}

Coxackievirus, responsible for the so-called hand, foot and mouth diease (HFMD), is highly infectious and primarily targets children (about 70\% of the patients are under 18 years old). It is a seasonal disease that often peaks in the summer (3rd quarter) and declines in the 4th quarter.  However, Figure \ref{fig:coxs_trends} shows that in the population of the case study, there is a consist climbing trend of the prevalance of Coxackie throughout the year of 2016.  Consequently, as seen in Figure \ref{fig:coxs_impact}, the total cost of Coxsackie and ECHO Infections increased 371\% in 2016. As expectd, this cost change is mainly due to the change in prevalence of Coxsackie the rate of which is as high as 186\% in 2016. A comparison of the demographics of 2015 and 2016 indicates that the increase of prevalence in 2016 is not due to any change in demographic distribution of patients (e.g., higher juvenile enrollment).  

Children are most vulnerable to coxsackievirus and places such as daycares and schools are at high-risk. There is no vaccine to prevent coxsackievirus infection. When a coxsackie outbreak occurs, daycares and schools will be closed and deep cleaning/santinizing of the entire facility is conducted. Such necessary actions can be quickly taken to prevent the future spread of the disease, if we can detect the emerging of the coxsackievirus outbreak at an early stage.  In this study, we applied our detection method at the end of each year but the same analysis can be performed on a monthly basis for early warnings of coxsackievirus outbreak. 
 
\newpage
{\bf System Improvement}

System improvement category identifies inpatient and outpatient emerging cost drivers that provide opportunities for the improvement of the healthcare provider system. As an example, outpatient chemotherapy drug administration is a procedure of infusing a chemotherapy specialty drug up to 1 hour. Multiple claims can be submitted for each additional hour.  

\begin{figure}[h!]
	\centering
	\includegraphics[width=\linewidth]{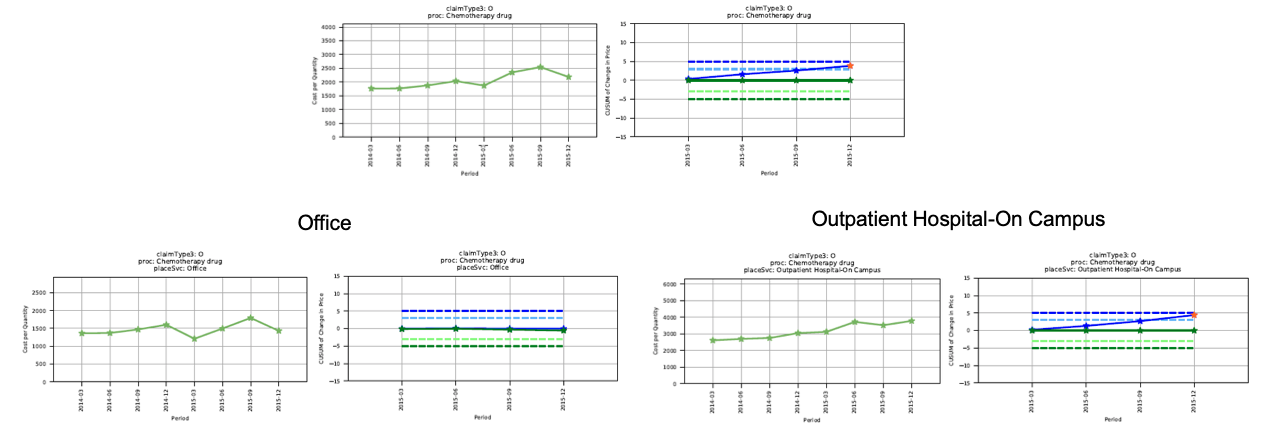}	
	\caption{Trends and CUMUM plots for chemotherapy drug administration cost at office versus hopital in 2015}	
	\label{fig:chemo_place}	
\end{figure}

The cost of this procedure increased 
about 47\% over the last four years 
in our analysis. Independent research study shows that the cost of chemotherapy drug administration is different by place of services\cite{hayes2015}. In particular, patients receiving chemotherapy treatment in the community oncology clinic had a 20\% to 39\% lower mean PMPM cost of care compared with those receiving chemotherapy in the hospital outpatient setting \cite{hayes2015}. This is consistent with our findings. In our study, we run our approach on all 700 qualified outpatient procedure group and place of service combinations yearly from 2013 to 2016. As seen in Figure \ref{fig:chemo_place}, the cost of chemotherapy drug administration received at on-campus facilities of a hospital almost doubles the cost of getting the same treatment in an office setting. In addition, the cost increase of chemotherapy drug administration is mainly attributed to outpatient hospital-on campus rather than office.  Table \ref{tab:chem_placetable} shows that in 2015 the total cost for chemotherapy patients received in hospital increased 40.8\%, which ranked as the highest in impact among around 700 qualified outpatient procedure group and place of service combinations in our study.  On the other hand, those reeived in office dropped 13.8\%.  According to our cost impact decomposition analysis, this cost increase in hospital is mainly due to an increase in price rather than prevalence.  If more chemotherapy services can be shifted to the office setting, the savings potential would be significant.

\begin{sidewaystable}
	\begin{tabular}{|c|c|c|c|c|c|c|c|c|c|c|c|c|c|}
		\hline
		\multirow{3}{*}{Glumetza}&\multicolumn{2}{c|}{\multirow{2}{*}{\makecell{Cost Among Pharmacy\\ Product Names}}}&\multicolumn{3}{c|}{\multirow{2}{*}{Change of Total Cost}}&\multicolumn{8}{c|}{Change of Contributing Factors(Trend \& Confidence)}\\
		\cline{7-14}
		&\multicolumn{2}{c|}{}&\multicolumn{3}{c|}{}&\multicolumn{4}{c|}{Price: Cost per Quantity of Services}&\multicolumn{4}{c|}{Use: Quantity of Services per Member}\\
		\cline{2-14}
		&\%&Rank&Impact&Rank&Rate&Impact&Rank&Rate&T\&C&Impact&Rank&Rate&T\&C\\
		\hline
		12/2013&0.02&200&\$0.01&209&11.9\%&\$0.02&108&25.7\%&$\uparrow$S&-\$0.01&-183&-10.1\%&$\updownarrow$N\\
		12/2014&0.02&190&\$0.02&179&18.0\%&\$0.01&181&12.4\%&$\uparrow$S&\$0.01&238&4.36\%&$\updownarrow$N\\
		12/2015&0.06&75&\$0.28&75&262\%&\$0.30&3&369\%&$\uparrow$VS&-\$0.02&-90&-17.3\%&$\downarrow$M\\
		12/2016&0.06&71&-\$0.25&-8&-63.0\%&\$0.24&4&95.7\%&$\downarrow$S&-\$0.49&-4&-82.0\%&$\downarrow$S\\
		\hline
	\end{tabular}
	\caption{Impact of change of Glumetza from 2013 to 2016. Confidence Key: N = no, M = moderate, S = strong, VS = very strong. Note that ``Rank'' is ranking among approximately 850 valid pharmacy drug product names.}
	\label{tab:glumetza_impact}
\vspace{5\baselineskip}
	\begin{tabular}{|c|c|c|c|c|c|c|c|c|c|c|c|c|c|c|}
		\hline
		\multicolumn{2}{|c|}{\multirow{3}{*}{\makecell{Procedure Group: \\Chemotherapy drug}}}&\multicolumn{2}{c|}{\multirow{2}{*}{Cost Among OP Claims}}&\multicolumn{3}{c|}{\multirow{2}{*}{Change of Total Cost}}&\multicolumn{8}{c|}{Change of Contributing Factors(Trend \& Confidence)}\\
		\cline{8-15}
		\multicolumn{2}{|c|}{}&\multicolumn{2}{c|}{}&\multicolumn{3}{c|}{}&\multicolumn{4}{c|}{Price: Cost per Quantity of Services}&\multicolumn{4}{c|}{Use: Quantity of Services per Member}\\
		\cline{3-15}
		\multicolumn{2}{|c|}{} &\%&Rank&Impact&Rank&Rate&Impact&Rank&Rate&T\&C&Impact&Rank&Rate&T\&C\\
		\hline
		\multirow{3}{*}{12/2015}&All$^{1}$&1.27&5&\$0.70&4&14.4\%&\$0.96&1&20.2\%&$\uparrow$M&-\$0.26&-4&-4.52\%&$\updownarrow$N\\
		                                      &Office$^{2}$&0.55&10&-\$0.34&-4&-13.8\%&\$0.07&30&2.18\%&$\updownarrow$N&-\$0.41&-1&-14.8\%&$\downarrow$M\\
		                                  &Hospital$^{2}$&0.72&6&\$1.00&1&40.8\%&\$0.73&1&27.3\%&$\uparrow$M&\$0.27&8&10.7\%&$\updownarrow$N\\
		\hline
	\end{tabular}
	\caption{Place of service comparisons of the impact of change fo outpatient chemotherapy drug administration in 2015. Confidence Key: N = no, M = moderate, S = strong, VS = very strong. Note that ``1'' is ranking among approximatedly 450 qualified outpatient procedure groups, and ``2'' is ranking among approximately 700 qualified outpatient procedure groups.}
	\label{tab:chem_placetable}
\end{sidewaystable}


\section*{Conclusion}
In this paper, we present a systematic approach to survelliance and detection of cost change drivers within large-scale healthcare claims data using hierarchical search strategies. Changes are detected based on year-to-year comparison of de-seasonalized and self-censored KPI change rates. An enhanced SPC algorithm is employed to automatically account for multiple change points (e.g., non-restarting CUSUM). Detection thresholds are learned via simulation or statistical theory to account for sampling errors and possible serial dependence of change rates. The impact of cost change is decomposed into multiple contributing factors (e.g., price, use, and prevalence) to help identify the root causes of the change. Furthermore, the detected change patterns could be used to identify ``utilization offsets'' of comparable treatment options and to quantify their cost impacts. When executed frequently, our approach provides data scientists and payers an early-warning alert system to surface emerging cost drivers and support early intervention for healthcare cost management. 

We demostrate our approach through a case study on IBM Waston Health MarketScan database. The cost drivers  are classified into muliple categories based on the type of potential actions and examples of detailed analysis of the cost drivers within each category are presented. 


\makeatletter
\renewcommand{\@biblabel}[1]{\hfill #1.}
\makeatother

\bibliographystyle{unsrt}

\end{document}